\documentclass[prd,a4paper,twocolumn]{revtex4}
\usepackage{graphicx}
\usepackage[]{subfigure}
\usepackage{epsfig}

\newcommand{\nc}{\newcommand}

\nc{\be}[1]{\begin{equation}\mbox{$\label{#1}$}}
\nc{\bea}[1]{\begin{eqnarray} \mbox{$\label{#1}$}}
\nc{\Section}[2]{\section{#2}\label{#1}}
\nc{\Bibitem}[1]{\bibitem{#1}}
\nc{\Label}[1]{\label{#1}}

\nc{\eea}{\end{eqnarray}}
\nc{\ee}{\end{equation}}

\nc{\bdm}{\begin{displaymath}}
\nc{\edm}{\end{displaymath}}
\nc{\dpsty}{\displaystyle}
\nc{\bc}{\begin{center}}
\nc{\ec}{\end{center}}
\nc{\ba}{\begin{array}}
\nc{\ea}{\end{array}}
\nc{\bab}{\begin{abstract}}
\nc{\eab}{\end{abstract}}
\nc{\btab}{\begin{tabular}}
\nc{\etab}{\end{tabular}}
\nc{\bit}{\begin{itemize}}
\nc{\eit}{\end{itemize}}
\nc{\ben}{\begin{enumerate}}
\nc{\een}{\end{enumerate}}
\nc{\bfig}{\begin{figure}}
\nc{\efig}{\end{figure}}

\nc{\arreq}{&\!=\!&}
\nc{\arrmi}{&\!-\!&}
\nc{\arrpl}{&\!+\!&}
\nc{\arrap}{&\!\!\!\approx\!\!\!&}
\nc{\non}{\nonumber}
\nc{\align}{\!\!\!\!\!\!\!\!&&}

\def\lsim{\; \raise0.3ex\hbox{$<$\kern-0.75em
      \raise-1.1ex\hbox{$\sim$}}\; }
\def\gsim{\; \raise0.3ex\hbox{$>$\kern-0.75em
      \raise-1.1ex\hbox{$\sim$}}\; }

\nc{\DOT}{\hspace{-0.08in}{\bf .}\hspace{0.1in}}
\nc{\Laada}{\hbox {$\sqcap$ \kern -1em $\sqcup$}}
\nc\loota{{\scriptstyle\sqcap\kern-0.55em\hbox{$\scriptstyle\sqcup$}}}
\nc\Loota{{\sqcap\kern-0.65em\hbox{$\sqcup$}}}
\nc\laada{\Loota}
\nc{\qed}{\hskip 3em \hbox{\BOX} \vskip 2ex}

\nc{\real}{{\rm I \! R}}
\nc{\Z}{{\sf Z \!\!\! Z}}
\nc{\complex}{{\rm C\!\!\! {\sf I}\,\,}}
\def\bigid{\leavevmode\hbox{\small1\kern-3.8pt\normalsize1}}
\def\id{\leavevmode\hbox{\small1\kern-3.3pt\normalsize1}}
\nc{\slask}{\!\!\!/}
\nc{\bis}{{\prime\prime}}
\nc{\pa}{\partial}
\nc{\na}{\nabla}
\nc{\ra}{\rangle}
\nc{\la}{\langle}
\nc{\goto}{\rightarrow}
\nc{\swap}{\leftrightarrow}

\nc{\EE}[1]{ \mbox{$\cdot10^{#1}$} }
\nc{\abs}[1]{\left|#1\right|}
\nc{\at}[2]{\left.#1\right|_{#2}}
\nc{\norm}[1]{\|#1\|}
\nc{\abscut}[2]{\Abs{#1}_{\scriptscriptstyle#2}}
\nc{\vek}[1]{{\rm\bf #1}}
\nc{\integral}[2]{\int\limits_{#1}^{#2}}
\nc{\inv}[1]{\frac{1}{#1}}
\nc{\dd}[2]{{{\partial #1}\over{\partial #2}}}
\nc{\ddd}[2]{{{{\partial}^2 #1}\over{\partial {#2}^2}}}
\nc{\dddd}[3]{{{{\partial}^2 #1}\over
    {\partial #2 \partial #3}}}
\nc{\dder}[2]{{{d #1}\over{d #2}}}
\nc{\ddder}[2]{{{d^2 #1}\over{d {#2}^2}}}
\nc{\dddder}[3]{{d^2 #1}\over
    {d #2 d #3}}
\nc{\dx}[1]{d\,^{#1}x}
\nc{\dy}[1]{d\,^{#1}y}
\nc{\dz}[1]{d\,^{#1}z}
\nc{\dl}[1]{\frac{d\,^{#1}l}{(2\pi)^{#1}}}
\nc{\dk}[1]{\frac{d\,^{#1}k}{(2\pi)^{#1}}}
\nc{\dq}[1]{\frac{d\,^{#1}q}{(2\pi)^{#1}}}

\nc{\bfT}{{\bf T }}

\nc{\cA}{{\cal A}}
\nc{\cB}{{\cal B}}
\nc{\cD}{{\cal D}}
\nc{\cE}{{\cal E}}
\nc{\cG}{{\cal G}}
\nc{\cH}{{\cal H}}
\nc{\cL}{{\cal L}}
\nc{\cO}{{\cal O}}
\nc{\cT}{{\cal T}}
\nc{\cN}{{\cal N}}
\nc{\cR}{{\cal R}}

\nc{\rvac}[1]{|{\cal O}#1\rangle}
\nc{\lvac}[1]{\langle{\cal O}#1|}
\nc{\rvacb}[1]{|{\cal O}_\beta #1\rangle}
\nc{\lvacb}[1]{\langle{\cal O}_\beta #1 |}
\nc{\bb}{\bar{\beta}}
\nc{\bt}{\tilde{\beta}}
\nc{\ctH}{\tilde{\cal H}}
\nc{\chH}{\hat{\cal H}}
\nc{\1}{\aa}
\nc{\2}{\"{a}}
\nc{\3}{\"{o}}
\nc{\4}{\AA}
\nc{\5}{\"{A}}
\nc{\6}{\"{O}}
\nc{\al}{\alpha}
\nc{\g}{\gamma}
\nc{\Del}{\Delta}
\nc{\e}{\textrm{e}}
\nc{\eps}{\epsilon}
\nc{\lam}{\lambda}
\nc{\Om}{\Omega}
\nc{\ve}{\varepsilon}
\nc{\mn}{{\mu\nu}}
\nc{\vp}{\varphi}

\nc{\rf}[1]{(\ref{#1})}
\nc{\nn}{\nonumber \\*}
\nc{\bfB}{\bf{B}}
\nc{\bfv}{\bf{v}}
\nc{\bfx}{\bf{x}}
\nc{\bfy}{\bf{y}}
\nc{\vx}{\vec{x}}
\nc{\vy}{\vec{y}}
\nc{\oB}{\overline{B}}
\nc{\oI}{\overline{I}}
\nc{\oR}{\overline{R}}
\nc{\rar}{\rightarrow}
\nc{\ti}{\times}
\nc{\slsh}{\hskip-5pt/}
\nc{\sm}{Standard~Model~}
\nc{\MP}{M_{\rm Pl}}
\nc{\mpl}{M_{\rm Pl}}
\nc{\tp}{t_{\rm Pl}}

\nc{\pmin}{p_{\rm min}}
\nc{\pmax}{p_{\rm max}}
\nc{\fo}{f_0}
\nc{\foi}{f_{0,i}\,}
\nc{\fop}{f_0^P}
\nc{\fou}{f_0^U}

\nc{\eff}{{\rm eff}}
\nc{\MT}{M_{\rm T}}
\nc{\ML}{M_{\rm L}}
\nc{\kk}{\vek{k}}
\nc{\pp}{{\rm p}}
\nc{\pt}{\partial_t}
\nc{\half}{{1\over 2}}
\nc{\w}{\omega}
\nc{\uhat}{\hat{U}_\w}

\nc{\etal}{\mbox{\it et al.}}
\nc{\ie}{{\it i.e. }}
\nc{\eg}{{\it e.g. }}
\nc{\trh}{T_{\rm RH}}
\nc{\ad}{{a'\over a}}
\nc{\bd}{{b'\over b}}
\nc{\Rd}{{R'\over R}}
\nc{\diag}{{\textrm{diag}}}
\nc{\mato}[1]{\tilde{#1}}
\nc{\sinn}{\textrm{sinn}}
\nc{\sech}{\textrm{sech}}
\nc{\I}{\textrm{I}}
\nc{\II}{\textrm{II}}
\nc{\III}{\textrm{III}}
\nc{\vev}[1]{\langle #1 \rangle}
\nc{\hyp}{\,\; F_{1{\hskip -16pt}2}{\hskip 11pt}}
\nc{\brhom}{\overline{\rho}_M}
\nc{\brho}{\overline{\rho}}
\nc{\rhob}{\overline{\rho}}
\nc{\Pb}{\overline{P}}
\nc{\bH}{\overline{H}}
\nc{\ep}{{1+4\eps}}

\nc{\deriv}[2]{ 
\frac{\mathrm{d}#1}{\mathrm{d}#2}
}

\def\smiley{\hbox{\large$\bigcirc$\hspace{-.80em}%
\raise.2ex\hbox{$\cdot\cdot$}\kern-.61em    %--- .56
\lower.2ex\hbox{\scriptsize$\smile$}}\ }

\def\frowney{\hbox{\large$\bigcirc$\hspace{-.80em}%
\raise.2ex\hbox{$\cdot\cdot$}\kern-.635em
\lower.2ex\hbox{\scriptsize$\frown$}}\ }

\begin{document}

\title{Revised WMAP constraints on neutrino masses and other
extensions of the minimal $\Lambda$CDM model}
\author{Jostein R. Kristiansen}
\email{j.r.kristiansen@astro.uio.no}
\affiliation{Institute of Theoretical Astrophysics, University of
  Oslo, Box 1029, 0315 Oslo, NORWAY}
\author{Hans Kristian Eriksen}
\email{h.k.k.eriksen@astro.uio.no}
\affiliation{Institute of Theoretical Astrophysics, University of
  Oslo, Box 1029, 0315 Oslo, NORWAY}
\affiliation{Centre of Mathematics for Applications, University of
  Oslo, Box 1053 Blindern, 0316 Oslo, NORWAY}
\affiliation{Jet Propulsion Laboratory, 4800 Oak Grove Drive, Pasadena
CA 91109, USA}
\affiliation{California Institute of Technology, Pasadena, CA 91125, USA}
\author{{\O}ystein Elgar\o y}
\email{oelgaroy@astro.uio.no}
\affiliation{Institute of Theoretical Astrophysics, University of
  Oslo, Box 1029, 0315 Oslo, NORWAY}
\date{\today}

\begin{abstract}

  Recently, two issues concerning the three-year Wilkinson Microwave
  Anisotropy Probe (WMAP) likelihood code
  were pointed out. On large angular scales ($l \lesssim 30$), a
  sub-optimal likelihood approximation resulted in a small power
  excess. On small angular scales ($l \gtrsim 300$), over-subtraction
  of unresolved point sources produced a small power deficit.  For a
  minimal six-parameter cosmological model, these two effects
  conspired to decrease the value of $n_s$ by $\sim 0.7 \sigma$.  In
  this paper, we study the change in preferred parameter ranges for
  extended cosmological models, including running of $n_s$, massive
  neutrinos, curvature, and the equation of state for dark energy. We
  also include large-scale structure and supernova data in our
  analysis. We find that the parameter ranges for $\alpha_s$,
  $\Omega_k$ and $w$ are not much altered by the modified analysis.
  For massive neutrinos the upper limit on the sum of the neutrino
  masses decreases from $M_\nu < 1.90$eV to $M_\nu < 1.57$eV when
  using the modified WMAP code and WMAP data only. We also find that
  the shift of $n_s$ to higher values is quite robust to these
  extensions of the minimal cosmological model.
\end{abstract}

\maketitle

\section{Introduction}   

The temperature fluctuations in the cosmic microwave background (CMB)
radiation have proved to be the single most important cosmological
observable we have today, and the high-precision full-sky maps
provided by the Wilkinson Microwave Anisotropy Probe (WMAP) play a
very important role in the determination of cosmological parameters
and preferred cosmological models \cite{spergel:2006}. One of the most
important conclusions both from the WMAP data alone, and from joint
analyses including other cosmological observables, is that a simple
six-parameter flat $\Lambda$CDM model fits the data very well, and
that extended models with additional free parameters do not improve
the fit significantly.

Because of the great impact of WMAP data, the 3-year data analysis
from the WMAP team has been subject to exhaustive cross-checking. In
particular, in refs.\ \cite{eriksen:2006} and \cite{huffenberger:2006}
two noticeable issues with the likelihood code as first presented by
the WMAP team were pointed out. First, the likelihood approximation
used between $13 \le l \lesssim 30$ appears to be inadequate,
effectively resulting in a $\sim 5\%$ power excess in this range
compared to an exact treatment. Second, the amplitude for the
unresolved point source spectrum used by the WMAP team was found to
over-subtract the actual contribution in the data, leading to a power
deficit at high $l$'s.

In \cite{huffenberger:2006}, the effect of these discrepancies were
studied for a minimal six-parameter cosmological model. This was done
both for WMAP data only, and with additional CMB data from small-scale
experiments. Their main finding was an increase in $n_s$, lowering the
significance of $n_s \neq 1$ from $\sim 2.7\sigma$ to $\sim 2.0
\sigma$.

In this paper we consider the effect on extended cosmological
models. We study how the modified WMAP likelihood affects the
preferred ranges of the running of the scalar spectral index, the
cosmological neutrino mass limits, spatial curvature, and the equation
of state for dark energy. We have also taken into account large-scale
structure (LSS) and type 1a supernovae (SNIa) data sets, to see
whether the shifts in preferred parameter ranges survive a more
thorough cosmological analysis. Further, we consider whether the shift
of $n_s$ to larger values is robust to changes in cosmological models
and data sets.

In the next section, we review both the methods and data we use. In
Section \ref{sec:results}, we report and comment upon our results,
before summarizing and concluding in Section \ref{sec:conclusion}.

\section{Data and methods}

\subsection{WMAP data} 
\label{sec:cmbanalysis}

The WMAP experiment is a NASA-funded satellite mission designed to
measure the CMB temperature anisotropies over the full sky at five
frequencies between 23 and 94 GHz with unprecedented angular
resolution and sensitivity. These measurements allow for an accurate
determination of the angular CMB power spectrum for angular scales
between, say, $l=2$ and 800 with three years of observations.

Estimation of this spectrum and the corresponding likelihood function
is a multi-step process. First, sky maps are generated from the raw
satellite data, and the instrumental noise is estimated. Second,
contaminants in the form of galactic and extra-galactic foregrounds
are removed from the sky maps, and severely contaminated regions are
removed completely from further analysis. Third, the power spectrum is
estimated with some algorithm, usually trading off computational
efficiency against accuracy. Fourth, a connection is made between the
power spectrum and the likelihood.

These steps are all described in detail for the three-year WMAP data
release in ref.\ \cite{hinshaw:2006}. The main result of these efforts
is a user-friendly Fortran 90 code that for an input power spectrum 
outputs the corresponding likelihood value. 
In principle, this piece of code may be
used as a ``black box''.

However, some care is warranted. In particular, two points were noted
in ref.\ \cite{eriksen:2006}. First, there is a $\sim 5\%$ descrepancy
between the temperature likelihood approximation used by the WMAP team
and an exact evaluation for $l \lesssim 30$. Second, there is a
$\sim 60\,\mu\textrm{K}^2$ discrepancy between the two spectra
observed at 61 and 94 GHz. The former is primarily due to estimator
approximations and secondarily to residual foregrounds. The latter
issue was later partly explained in terms of an excessive point source
correction applied to the WMAP spectrum \cite{huffenberger:2006}.

In the present paper, we therefore use two versions of the WMAP
likelihood. The first version is simply the official code as provided
on LAMBDA\footnote{http://lambda.gsfc.nasa.gov; version v2p1.}. The
second version includes two modifications to this code: At $l \le 30$,
we replace both the WMAP pixel-based likelihood and the
pseudo-$C_{l}$-based likelihood with an exact Gibbs sampling based
estimator \cite{eriksen:2006}. Then the spectrum amplitude of
unresolved point sources (relative to 41 GHz) is adjusted from $A =
0.017 \,\mu\textrm{K}^2 \textrm{sr}$ to $A = 0.011 \,\mu\textrm{K}^2
\textrm{sr}$ \cite{huffenberger:2006}. We do not marginalize over the
SZ (Sunyaev-Zeldovich) amplitude in our analyses.

\subsection{Other data sets used}

In our analysis we use additional CMB data, data from LSS surveys,
SNIa data and additional priors on the Hubble parameter and baryon
content of the universe.

\subsubsection{Other CMB observations}

To probe a larger range of angular scales in the CMB power spectrum we
use CMB data from ACBAR \cite{kuo:2004} and BOOMERanG
\cite{jones:2005, piacentini:2005, montroy:2005}.

\subsubsection{Large scale structure}

Large scale structure surveys probe the matter distribution in the
universe by measuring the galaxy-galaxy power spectrum $P_g(k,z) =
\langle |\delta_{g}(k,z)|^2 \rangle$. In the linear perturbation
regime it is expected that this galaxy-galaxy spectrum is proportional
to the total matter power spectrum, $P_m$, through the simple relation
$P_g = b^2 P_m$, where $b$ is called the bias parameter.

There are two galaxy surveys of comparable size, namely the 2 degree
Field Galaxy Redshift Survey (2dF) \cite{colless:2003} and the Sloan
Digital Sky Survey (SDSS) \cite{tegmark:2004}. In our analysis we use
data from both these surveys.

\subsubsection{Type 1a supernovae}

Probing the luminosity-redshift relation of SNIa is one of the most
direct measurements of cosmological expansion, and thus one of the
most powerful pieces of evidence for the existence of dark energy. In
our analysis we use SN1a data from the Supernova Legacy Survey (SNLS)
\cite{astier:2006}, which is a dedicated SNIa survey currently
including 71 SNIa in the redshift range $z=0.2-1$.

\subsubsection{Additional priors}

In addition to the CMB, LSS and SNIa data sets mentioned above, we
impose priors on the Hubble parameter, the baryon content in the
universe, and the position of the LSS baryonic peak.

From the Hubble Space Telescope Key Project (HST) we have adopted a
prior on the Hubble parameter of $h=0.72 \pm 0.08$
\cite{freedman:2001}. The constraint on the baryon density today was
chosen to be $\Omega_b h^2 = 0.022 \pm 0.002$ from Big Bang
nucleosynthesis (BBN) \cite{burles:2001, cyburt:2004}.

From the detection of baryonic acoustic oscillations (BAO) in the
sample of luminous red galaxies (LRG) in the SDSS survey
\cite{eisenstein:2005} it is also possible to put a constraint on the
combination of parameters
\begin{equation}
A_{\textrm{BAO}} \equiv \left[ D_M(z)^2 \frac{z}{H(z)} \right]^{1/3}
\frac{\sqrt{\Omega_m H_0^2}}{z},
\end{equation}
where $D_M(z)$ is the comoving angular diameter distance. The BAO
constraint can then be written as
\begin{equation} \label{eq:A}
A_{\textrm{BAO}} = 0.469 \left( \frac{n_s}{0.98} \right)^{-0.35} (1+0.94 f_\nu) \pm 0.017,
\end{equation}
where the fit to the neutrino fraction, $f_\nu = \Omega_\nu/\Omega_m$
is given by \cite{goobar:2006}. For $z$ we adopt the redshift of a
typical LRG in the SDSS sample, $z=0.35$. For models with non-zero
$\alpha_s$, we substitute $n_s$ in eq. (\ref{eq:A}) with an effective
$n_s$ given by
\begin{equation}
n_{s,\textrm{eff}}(k_1) \equiv \left. \dder{\ln P}{\ln k} \right|_{k=k_1} + 1 = n_s(k_0) + \alpha_s \ln(k_1/k_0), 
\end{equation}
where $P$ is the primordial power spectrum given by
\begin{equation} \label{eq:Pdef}
\ln P = \ln A_s + (n_s -1) \ln(k/k_0) + \alpha_s/2 \ln(k/k_0)^2,
\end{equation}
and $k_0$ is set to $0.05\textrm{Mpc}^{-1}$. We use $k_1 = 0.01
h\textrm{Mpc}^{-1}$, which is approximately the scale associated with
the baryonic peak in the LRG sample. Our results are robust to changes
of $k_1$ around this value, as the preferred values of $\alpha_s$ are
small in the models considered here.

\subsection{Parameter estimation}

The parameter estimation process is based on the publicly available
CosmoMC code \cite{lewis:2002}, using the data and likelihoods
described above.

For each model, we compute the corresponding parameter confidence
intervals using three different combinations of data sets, named A, B
and C (see Table \ref{tab:datasets}). Data set A includes the
three-year WMAP data only; data set B includes also CMB data from
ACBAR and BOOMERanG; and data set C also LSS and SNIa data sets and
priors from HST, BBN and BAO.

As a basic six-parameter cosmological model we use the parameters
$\{\Omega_b h^2, \Omega_m, \log(10^{10}A_s), h, n_s, \tau\}$. The
exact parameter definitions are given by the CosmoMC code. $\Omega_b$
is the ratio of baryons to the total energy density; $h$ is the Hubble
parameter today; $\Omega_m$ is the ratio of matter to the total energy
density today; $A_s$ sets the amplitude of the primordial fluctuations; $n_s$
is the tilt of the primordial power spectrum; and $\tau$ is the
reionization optical depth.

We then extend the six-parameter model by adding the parameters
${\alpha_s, r, M_\nu, \Omega_k, w}$ one by one (except for $\alpha_s$
and $r$ which are added simultaneously). Here, $\alpha_s$ is the
running of $n_s$, defined in eq.(\ref{eq:Pdef}); $r$ is the ratio
of tensor to scalar fluctuations; $M_\nu$ is the sum of the neutrino
mass eigenstates, $M_\nu = \sum m_\nu = 93.14 \Omega_\nu h^2$;
$\Omega_k$ is the amount of spatial curvature; and $w$ is the equation
of state parameter for dark energy. Finally, we vary all 11 parameters
simultaneously.

For all combinations of data and parameter sets, we carry out a
similar analysis both with the standard WMAP likelihood code as
provided and with the two modifications described above. 

\begin{table}[htb]
\begin{tabular}{cl}
\hline \hline
Data set & Observations included \\
\hline
A & WMAP \\
B & WMAP + ACBAR + BOOMERanG \\
C & WMAP + ACBAR + BOOMERanG + SDSS \\
& + 2dF + SNLS + HST + BBN + BAO \\
\hline
\end{tabular}
\caption{The different combinations of data sets used in this analysis.}
\label{tab:datasets}
\end{table}

\section{Results} \label{sec:results}
 
\subsection{Minimal six-parameter model} \label{sec:sixpar}

We start with the simple six-parameter model having the free
parameters $\{\Omega_b h^2, \Omega_m, \log(10^{10}A_s), h, n_s,
\tau\}$. This was first done in ref. \cite{huffenberger:2006} for the
combinations of data sets A and B (see Table \ref{tab:datasets}). For
$n_s$ they reported a $\sim 0.7 \sigma$ shift to higher values when
applying the modified analysis. The other parameters were only subject
to small shifts of their mean values.

We repeat this analysis here, but also include data set C in the
analysis. Consistent with ref. \cite{huffenberger:2006} we find
that only $n_s$ is notably affected by the modified likelihood, and
the shift of $n_s$ also remains for data set C, in which
case the value for $n_s$ changes from $n_s = 0.961 \pm 0.014$ to $n_s
= 0.971 \pm 0.014$. This corresponds to a shift of $\sim 0.8 \sigma$,
and weakens the significance of $n_s \neq 1$ from $\sim 2.9$ to $\sim
2.1$ for this data set. As can be seen from
refs. \cite{eriksen:2006, huffenberger:2006}, the low-$l$ and
point-source corrections contribute almost equally much to the shift
in $n_s$. In ref. \cite{eriksen:2006} they found a mean value of $n_s$
of $0.961$ when applying only the low-$l$ corrections. 

That the shift of $n_s$ survives when adding
LSS data is not very surprising; $n_s$ is less sensitive to LSS than
to CMB data because of the larger dynamic range and higher precision
of the latter observations. The resulting values of $n_s$ are
summarized in Table \ref{tab:sixparameter}.

\begin{table}[htb]
\begin{tabular}{cccc}
\hline \hline
Data set & WMAP code &\qquad& Modified code \\
\hline
$A$ &  $0.954\pm0.016$    && $0.966\pm0.016$ \\
$B$ &  $0.958\pm0.016$    && $0.969\pm0.016$ \\
$C$ &  $0.961\pm0.014$    && $0.971\pm0.014$ \\
\hline
\end{tabular}
\caption{Results for $n_s$ in a six-parameter model. The values in the
second column are found using the WMAP likelihood code, while the
values in the third column are calculated using the modifications
described in section \ref{sec:cmbanalysis}. All errors are 1$\sigma$.} 
\label{tab:sixparameter}
\end{table}

\subsection{Running of spectral index}

Next we consider how the modified WMAP likelihood affects the
constraints on $\alpha_s$. The simplest inflationary models predict an
$\alpha_s$ that is slightly different from 0, and thus information on
$\alpha_s$ can provide us with valuable information on inflationary
mechanisms. Following \citet{spergel:2006}, we marginalize over the
ratio of tensor to scalar fluctuations, $r$, since models with
negative $\alpha_s$ often correspond to large tensor modes.

Our results are summarized in Table \ref{tab:alphas}. We find that the
likelihood modifications have no major effect on the constraints on
$\alpha_s$, but we observe a small increase of $\sim 0.2 \sigma$ in
the significance of $\alpha_s \neq 0$.

\begin{table}[htb]
\begin{tabular}{cccc}
\hline \hline
Data set & WMAP code &\qquad& Modified code \\
\hline
$A$ & $-0.050\pm0.027$  &&  $-0.052\pm0.027$ \\
$B$ & $-0.052\pm0.026$ &&  $-0.056\pm0.025$\\
$C$ & $-0.013\pm0.020$ && $-0.014\pm0.019$\\
\hline
\end{tabular}
\caption{Estimated values for $\alpha_s$ from the WMAP likelihood code
  and our modified code. The other parameters, including $r$, are
  marginalized over.}
\label{tab:alphas}
\end{table}

\subsection{Massive neutrinos}

Another natural extension of the minimal six-parameter model is the
addition of massive neutrinos. This is motivated by observations of
neutrino oscillations, which show that neutrinos indeed are massive.

Because of their low mass, neutrinos act like a warm dark matter
component in the universe. Given the energy fraction of massive
neutrinos today, $\Omega_\nu$, one can easily find a limit on the sum
of the neutrino masses, $M_\nu$, by the relation $\Omega_\nu h^2 =
M_\nu/93.14 \textrm{eV}$. (See ref.\ \cite{elgaroy:2005} or
\cite{lesgourgues:2006} for a review of the cosmological properties of
massive neutrinos.)

At present, the best upper limits on the absolute mass scale of
neutrinos come from cosmology. The current cosmological 95\% C.L.\
limits range from $M_\nu < 0.17$eV \cite{seljak:2006}, relying on
extensive use of different data sets, to $M_\nu \lesssim 2.0$eV for
WMAP data only \cite{ichikawa:2005, spergel:2006, fukugita:2006}. In
ref.\ \cite{ichikawa:2005} they also pointed out that it will be
difficult to push the upper limit on $\Omega_\nu h^2$ much below
$\Omega_\nu h^2 \lesssim 0.017$ using CMB data only. This corresponds
to a neutrino mass limit $M_\nu \lesssim 1.5$eV. For smaller neutrino
masses, neutrinos will still be relativistic at the time of
recombination, and thus the effects of the neutrino masses will not be
fully revealed in the CMB power spectrum.

In our analysis we assume three species of massive neutrinos with
degenerate masses. The assumption of degenerate masses has been shown
to be very good for the mass regime that we are working in here
\cite{slosar:2006}. The resulting neutrino mass limits are summarized
in Table \ref{tab:Mnu}.  We see that when using WMAP data alone, the
upper limit on $M_\nu$ is significantly improved by the modified
analysis, and that we are approaching the limit of how tight
constraints on $M_\nu$ we can find from CMB data alone. If we analyze
the data applying only the low-$l$ corrections to the WMAP code, the neutrino mass
limit becomes $M_\nu < 1.69eV$, which shows that both the low-$l$ and
point-source corrections are important also for this model. 

\begin{figure}
\begin{center}
  \epsfig{figure=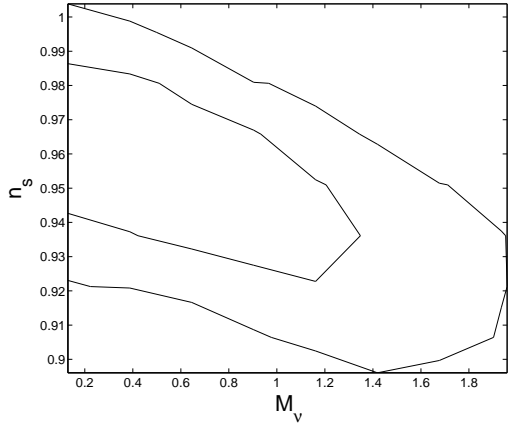,width=0.42\textwidth}
\caption{68\% and 95\% C.L. contours in the $M_\nu$-$n_s$ plane for
  the modified likelihood code and a seven-parameter model with free
  neutrino mass, using WMAP data only. There is a slight degeneracy between the two
  parameters, and larger values for $n_s$ will put tighter upper
  limits on $M_\nu$.}
\label{fig:Mnu_and_ns}
\end{center}
\end{figure}

Our improved $M_\nu$ limit can be understood by the slight degeneracy
between the $M_\nu$ and $n_s$ parameters, in that a larger value of
$n_s$ provides less space for a large $M_\nu$. This can be seen from
the contour plot in the $M_\nu$-$n_s$ plane in Figure
\ref{fig:Mnu_and_ns}. The degeneracy can be understood by the fact
that both $n_s$ and $M_\nu$ have impact on the small-scale behavior of the CMB power
spectrum.

By definition, $n_s$ sets the tilt of the primordial spectrum. If
$M_\nu$ is of order $\lesssim 2$eV, it will affect the power on scales
smaller than $l \sim 300$. This happens because the perturbations of
the gravitational potential on scales smaller than this are suppressed
by neutrino free streaming, which in turn boosts the acoustic
oscillations \cite{dodelson:1996}. As $M_\nu$ increases and more of
the dark matter consists of massive neutrinos, this boost of small
scale power also increases. Therefore, a large value of $n_s$
increases the power on small scales, leaving less room for $M_\nu$ to
add further power without coming in conflict with data. $M_\nu$ also
affects the heights of the peaks on larger scales, but this can to a
large extent be compensated for by adjusting the values of $\Omega_m$
and $\Omega_b h^2$.

It is interesting to notice that the upper limit on $M_\nu$ actually
weakens when we add small scale CMB data sets. From Table
\ref{tab:sixparameter} we also see that the preferred value of $n_s$
also increases when these data sets are included. This indicates that
the small-scale power from ACBAR and BOOMERanG is higher than what one
would expect from the WMAP data. When adding massive neutrinos to the
minimal six-parameter model this increment in small-scale power can be
partly accommodated by increasing $M_\nu$ instead of $n_s$. This will
increase the small-scale power without altering the fit to the large
scale spectrum.

We also see that when we add LSS and SN1a data, the neutrino mass
limit is no longer affected by the modified WMAP analysis, as the
additional constraints on $M_\nu$ are mainly determined by LSS data.
In the LSS power spectrum neutrino free-streaming will suppress small
scale power, and a larger $n_s$ will in this case allow for a
\emph{larger} $M_\nu$. This effect then cancels out the improved
$M_\nu$ limit found from the WMAP data.

\begin{table}[htb]
\begin{tabular}{cccc}
\hline \hline
Data set & WMAP code &\qquad & Modified code \\
\hline
$A$ & {1.90eV} && {1.57eV}\\
$B$ & {2.13eV} && {1.72eV}\\
$C$ & {0.45eV} && {0.45eV}\\
\hline
\end{tabular}
\caption{Estimated 95\% C.L. upper limits on $M_\nu$.}
\label{tab:Mnu}
\end{table}

\subsection{Spatial curvature}

Next we add spatial curvature, $\Omega_k$, to our six-parameter
model. However, altering the geometry of universe mainly affects the
positions of the CMB acoustic peaks, while the likelihood
modifications mostly concern the amplitude and tilt of the power
spectrum. A priori, one would therefore not expect any significant
changes in $\Omega_k$. And as seen in Table \ref{tab:Omk}, this is
indeed the case. For all data sets, $\Omega_k = 0$ is within $\sim 1
\sigma$, in good agreement with the results from the WMAP team
\cite{spergel:2006}.

The large improvement of the limits on $\Omega_k$ for data set C can
to a large extent be understood by the well-known degeneracy between
$\Omega_k$ and $h$, where negative values of $\Omega_k$ can be
accommodated by a small $h$. Therefore, when imposing the HST prior on
$h$, the allowed range of $\Omega_k$ is significantly constrained.

\begin{table}[htb]
\begin{tabular}{crlcrl}
\hline \hline
Data Set & \multicolumn{2}{c}{WMAP code} &\qquad& \multicolumn{2}{c}{Modified code} \\
\hline
$A$ &  $-0.057$ & $^{+0.050}_{-0.056}$   &&$-0.057$ & $^{+0.050}_{-0.057} $  \\
$B$ &  $-0.056$& $^{+0.052}_{-0.062}$ && $-0.055$&$^{+0.048}_{-0.055}$ \\
$C$ &  $-0.005$&$\pm0.007$ && $-0.006$&$\pm0.007$ \\
\hline
\end{tabular}
\caption{Estimated values for $\Omega_k$.}
\label{tab:Omk}
\end{table}

\subsection{Dark energy equation of state} \label{sec:w}

The nature of dark energy is one of the major questions in cosmology
today. In the minimal six-parameter model, the dark energy is assumed
to be a cosmological constant with $w = -1$, and this has been shown
to agree well with current cosmological data (see, e.g., ref.\
\cite{spergel:2006}). Here we test whether the modified WMAP
likelihood code alters the preferred values for $w$. In the following
analysis we assume that $w$ is independent of redshift.

The main effect of $w$ on the CMB power spectrum is to shift the
position of the acoustic peaks by altering the expansion history of
the universe. Therefore we would not expect the limits on $w$ to be
much affected by the new WMAP likelihood analysis. Still, from Table
\ref{tab:w} we notice a small shift of order $\sim 0.2 \sigma$ to
smaller values of $w$ when using CMB data only. This happens because a
smaller $w$ will enhance the late integrated Sachs-Wolfe effect, which
results in a suppression of large-scale fluctuations in the observed
CMB power spectrum. The slightly smaller preferred value of $w$ in the
modified analysis is accompanied by small changes also in $h$ and
$\Omega_m$ to shift the peaks back in position. As more data is added,
we see that the modified analysis has no effect on $w$ anymore, and
that a cosmological constant still fits the data.

\begin{table}[htb]
\begin{tabular}{cccc}
\hline \hline
Data set & WMAP code &\qquad& Modified code \\
\hline
$A$ & $-0.98\pm0.41$ && $-1.05\pm0.39$\\
$B$ &    $-0.97\pm0.41$&&  $-1.03\pm0.37$\\
$C$ &  $-1.00\pm0.07$ && $-1.00\pm0.07$\\
\hline
\end{tabular}
\caption{Estimated values for $w$.}
\label{tab:w}
\end{table}

\subsection{11-parameter model}

Finally we vary all 11 parameters $\{\Omega_b h^2, \Omega_m,
\log(10^{10}A_s), h, n_s, \tau, \alpha_s, r, M_\nu, \Omega_k, w \}$
simultaneously. The results for the parameters $n_s$, $\alpha_s$,
$\Omega_k$, $w$ and $M_\nu$ are shown in Table \ref{tab:fullset}. Here
we see that all effects found in the more restricted models above are
also present in this extended model.

For $n_s$ the likelihood corrections still result in a mean value that
is $\sim 0.02$ larger than with the original likelihood. However, as
the uncertainty in $n_s$ is increased (mainly due to a degeneracy with
$\alpha_s$), this shift is not as statistically significant as it was
in the six-parameter model in subsection \ref{sec:sixpar}. Further,
for data set C we see that the preferred value of $n_s$ is in fact not
much changed by the modified analysis. Rather, the power spectrum
changes are accommodated by a slightly lower value of $\alpha_s$ to
account for the smaller power for low $l$'s in the CMB power spectrum.

For $\alpha_s$, there are no significant changes of the preferred
values by the 11-parameter model, and we see that $\alpha_s=0$ is
still consistent with the data.

The same is the case for $\Omega_k$. Here the modified analysis shifts
the preferred values to slightly more negative values, but with all
data sets, $\Omega_k = 0$ remains well within $1 \sigma$ of its mean
value.

Also for $M_\nu$ we find the same effects as above. When using CMB
data only, the upper limit on $M_\nu$ is improved by the new WMAP
likelihood analysis. As expected, the limit becomes weaker for the
extended parameter set, but from WMAP data alone we still find a
$M_\nu$ limit that is better than what is found in earlier papers,
even with this large parameter space. Also, we see that the $M_\nu$
limit still becomes weaker by adding the ACBAR and BOOMERanG data
sets.

For the dark energy equation of state, we still find that the modified
analysis prefer slightly lower values for $w$. In this extended model
the preferred values of $w$ are shifted to lower values than in the
more restricted model in subsection \ref{sec:w}, which is mainly
accommodated by the preferred negative value of $\Omega_k$ (as $w$ and
$\Omega_k$ shifts the positions of the acoustic peaks in opposite
directions). Still $w=-1$ remains within $1 \sigma$ for all of the
data sets.

\begin{figure}[htb]
\begin{center}
\epsfig{figure=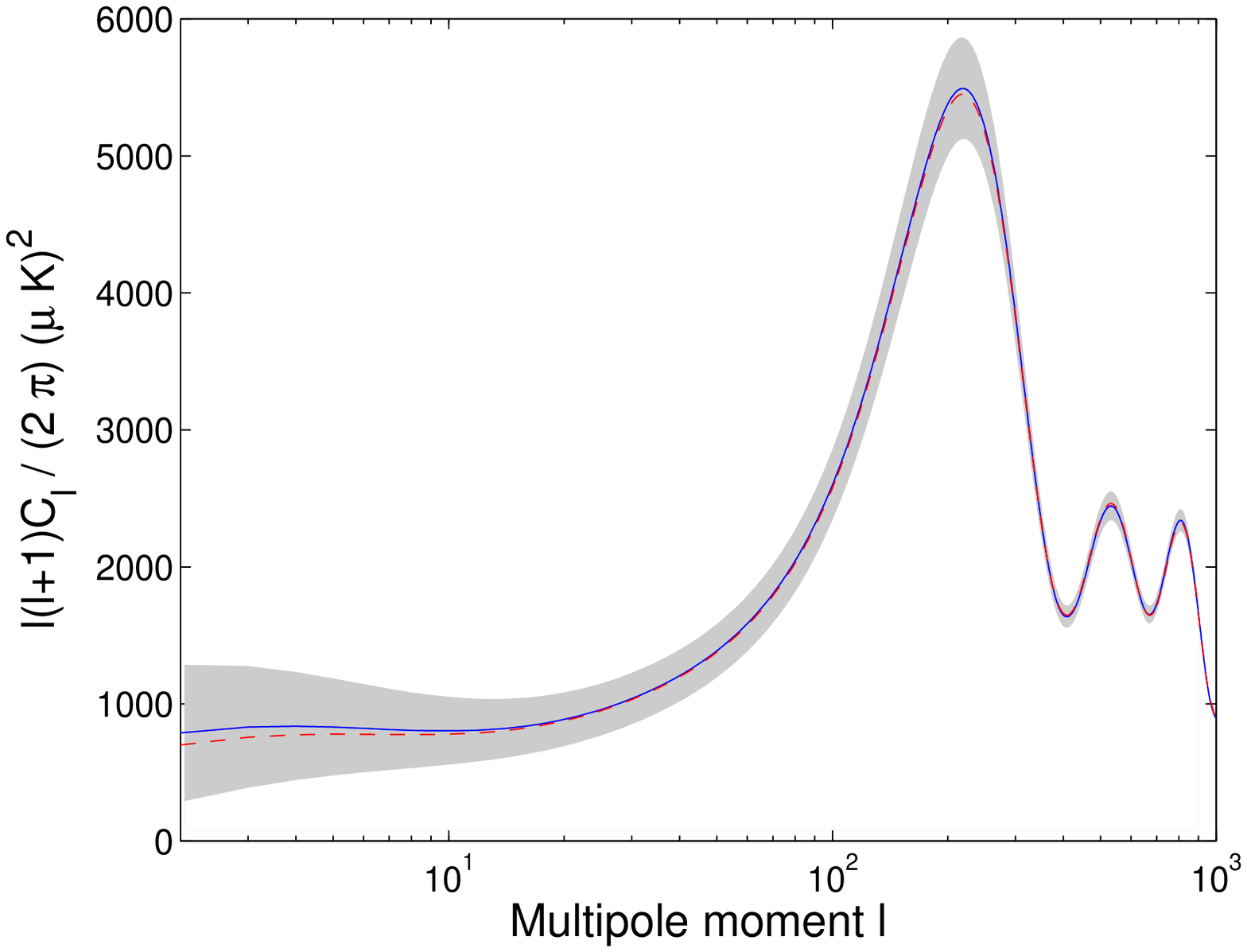,width=0.42\textwidth}
\epsfig{figure=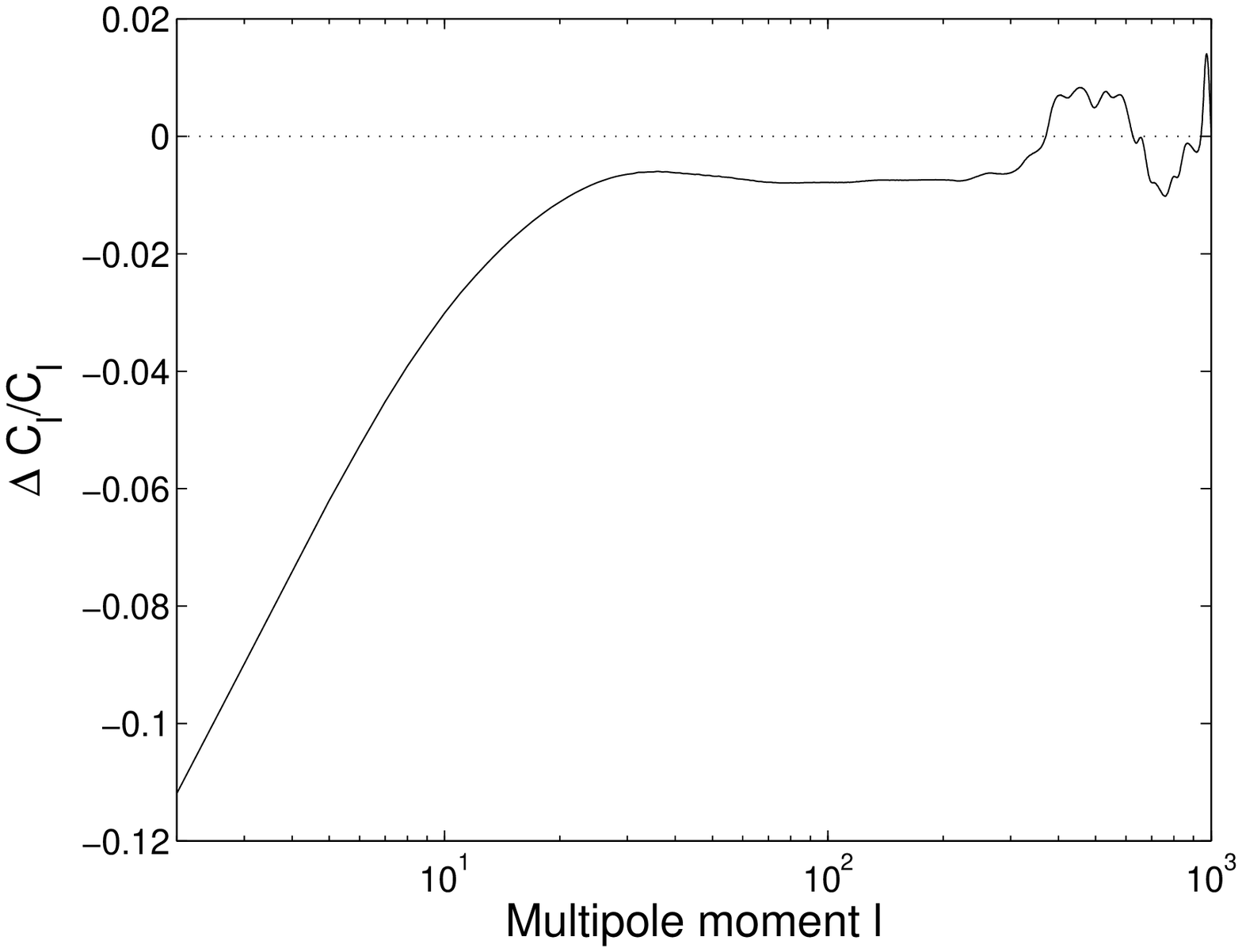,width=0.42\textwidth}
\caption{Upper panel: The CMB power spectrum for the best-fit 11-parameter model with
  the WMAP likelihood code (solid blue line) and the low-$l$ and point
  source corrected likelihood code (dashed red line). The gray shading
  shows the cosmic variance around the blue line. Note the small
  discrepancies at both low and high $l$'s. Lower panel: The  relative
  difference between the two power spectra, $(C_l^{\textrm{modified}}-C_l^\textrm{WMAP})/C_l^\textrm{WMAP}$. }
\label{fig:powerspectrum}
\end{center}
\end{figure}

In Figure \ref{fig:powerspectrum}, we show the best-fit power spectrum
for the 11-parameter the model both from the analysis with the
standard WMAP likelihood code and the new point-source and low-$l$
corrected likelihood analysis. We see that the discrepancy is most
notable for $l \lesssim 100$.

\begin{table}[htb]
\begin{tabular}{crlcrl}
\hline \hline
Parameter & \multicolumn{2}{c}{WMAP code} &\qquad& \multicolumn{2}{c}{Modified code} \\
\hline
\hline
\multicolumn{6}{c}{Data set A} \\
\hline
$n_s(0.05)$ & $0.863$&$\pm0.047$ && $0.880$&$\pm0.046$ \\
$\alpha_s$ & $-0.051$&$\pm0.029$ && $-0.050$&$\pm0.029$\\
$\Omega_k$ & $-0.019$&$^{+0.052}_{-0.053}$ && $-0.027$&$^{+0.049}_{-0.053}$\\
$w$ & $-1.43$&$\pm1.09$ && $-1.53$&$\pm1.12$\\
$M_\nu$ & \multicolumn{2}{l}{$<$2.09eV @ 95\% C.L.} && \multicolumn{2}{l}{$<$1.66eV @ 95\% C.L.}\\
\hline
\multicolumn{6}{c}{Data set B}\\
\hline
$n_s(0.05)$ & $0.859$&$\pm0.042$ && $0.875$&$\pm0.041$ \\
$\alpha_s$ & $-0.055$&$\pm 0.027$ && $-0.055$&$\pm0.027$\\
$\Omega_k$ & $-0.010$&$^{+0.047}_{-0.050}$ && $-0.018$&$^{+0.048}_{-0.053}$\\
$w$ & $-1.33$&$\pm1.02$&& $-1.43$&$\pm1.10$\\
$M_\nu$ & \multicolumn{2}{l}{$<$2.33eV @ 95\% C.L.} && \multicolumn{2}{l}{$<$2.02eV @ 95\% C.L.}\\
\hline
\multicolumn{6}{c}{Data set C} \\
\hline
$n_s(0.05)$ &$0.954$&$\pm0.038$  &&  $0.954$&$\pm0.036$ \\
$\alpha_s$ & $-0.003$&$\pm0.028$ && $-0.012$&$\pm0.026$\\
$\Omega_k$ & $-0.001$&$\pm0.012$ && $-0.001$&$^{+0.011}_{-0.013}$\\
$w$ & $-1.05$&$\pm0.09 $&& $-1.05$&$\pm0.09$\\
$M_\nu$ & \multicolumn{2}{l}{$<$0.51 eV @ 95\% C.L.} && \multicolumn{2}{l}{$<$0.52eV @ 95\% C.L.}\\
\hline
\end{tabular}
\caption{Parameter results for the model that includes free
$\alpha_s$, $r$, $M_\nu$, $w$ and $\Omega_k$.}
\label{tab:fullset}
\end{table}

\section{Conclusion} \label{sec:conclusion}

In refs.\ \cite{eriksen:2006} and \cite{huffenberger:2006}, two
modifications to the three-year WMAP likelihood were presented.
Relative to the power spectrum presented by Hinshaw
\cite{hinshaw:2006}, they found a power deficit for low values of $l$
due to inaccurate likelihood approximation, and a small power excess
for high values of $l$ due to over-subtracted unresolved point
sources.

The impact on the inferred cosmological parameter intervals from these
corrections in a minimal six-parameter model was studied in ref.\
\cite{huffenberger:2006} using CMB data only. Their single most
important result was an increase in the preferred value of $n_s$,
lowering the significance of $n_s \neq -1$ from $\sim 2.7 \sigma$ to
$\sim 2.0 \sigma$. In the present paper, we have extended that
analysis to also account for cosmological models including a non-zero
running of $n_s$, massive neutrinos, curvature and $w \neq -1$. We
have also added LSS and SNIa data sets to our analysis to see if the
parameter shifts induced by the modified WMAP analysis survive when
adding more data sets.

We found that the shift of $n_s$ to larger values survives when we add
LSS and SNIa data. However, when we apply all data sets in the full
11-parameter model, $n_s$ is not affected much by the modified WMAP
analysis anymore. This is mainly due to the extra freedom with
$\alpha_s$.

For the extended models, we found that the preferred values of
$\alpha_s$, $\Omega_k$ and $w$ are not significantly affected by the
modified analysis. When including massive neutrinos we found that the
upper limit on $M_\nu$ when using WMAP data alone was reduced from
$M_\nu < 1.90$eV to $M_\nu < 1.57$eV. A similar improvement in the
$M_\nu$ limit could not be observed when adding LSS and SN1a data,
since the higher preferred value of $n_s$ will allow for larger
neutrino masses in the LSS power spectrum.

Since the initial publication of the two re-analysis papers, refs.\
\cite{eriksen:2006} and \cite{huffenberger:2006}, and the present
paper, the WMAP team has released a new version of their likelihood
code\footnote{Version v2p2p1} that implements the suggested low-$l$
correction and a revised point-source correction.  Using this updated
likelihood code, we find $n_s = 0.959 \pm 0.016$ for the six-parameter
model and the WMAP data only. Including massive neutrinos, this code
gives an upper bound of $M_\nu<1.75\textrm{eV}$ from WMAP data only.
The difference is due to the point source amplitude and corresponding
error adopted by the WMAP team, which do not match perfectly that of
ref.\ \cite{huffenberger:2006}. Unfortunately, full details on the
WMAP approach are not currently available, and final assessment of
this issue must therefore await the release of the revised WMAP3
papers.

To conclude, the modified analysis does not strengthen the case for
non-standard cosmological parameters, and the standard flat
$\Lambda$CDM model still provides an excellent fit to data.
   
\acknowledgments

JRK and {\O}E acknowledge support from the Research Council of
Norway through project numbers 159637 and 162830. HKE acknowledges financial
support from the Research Council of Norway.

%%%%%%%%%%%%%%%%%%%%%%%%%%%%%%%%%

%\bibliography{cites}

\end{document}